\documentclass[reprint,amsmath,amssymb,aps,floatfix,groupedaddress,superscriptaddress,nofootinbib]{revtex4-1}
\pdfoutput=1
\usepackage{float}
\usepackage{hyperref}% add hypertext capabilities
\usepackage{graphicx,color}% Include figure files
\usepackage{dcolumn}% Align table columns on decimal point
\usepackage{bm}% bold math
\usepackage{here}
\usepackage{comment}
\usepackage[paperwidth=210mm,paperheight=297mm,centering,hmargin=1.5cm,vmargin=2.0cm]{geometry}
\usepackage{tabularx}
\newcolumntype{Y}{&gt;{\centering\arraybackslash}X} %中央揃え
\hypersetup{
 setpagesize=false,
 bookmarksnumbered=true,%
 bookmarksopen=true,%
 colorlinks=true,%
 linkcolor=blue,
 citecolor=blue,
 urlcolor=blue,
}
\usepackage[left]{lineno}% Enable numbering of text and display math
\begin{document}

\preprint{APS/123-QED}

\title{\bf Importance of separated efficiencies between positively and negatively \\ charged particles for cumulant calculations}% Force line breaks with \\
%\thanks{A footnote to the article title}%

\author{Toshihiro Nonaka}
 \email{tnonaka@rcf.rhic.bnl.gov}
\affiliation{Center\,for\,Integrated\,Research\,in\,Fundamental\,Science\,and\,Engineering,\,University\,of\,Tsukuba,\,Tsukuba,\,Ibaraki\,305,\,Japan}
\author{Tetsuro Sugiura}
 \email{tsugiura@rcf.rhic.bnl.gov}
\affiliation{Center\,for\,Integrated\,Research\,in\,Fundamental\,Science\,and\,Engineering,\,University\,of\,Tsukuba,\,Tsukuba,\,Ibaraki\,305,\,Japan}
\author{ShinIchi Esumi}
\affiliation{Center\,for\,Integrated\,Research\,in\,Fundamental\,Science\,and\,Engineering,\,University\,of\,Tsukuba,\,Tsukuba,\,Ibaraki\,305,\,Japan}
\author{Hiroshi Masui}
\affiliation{Center\,for\,Integrated\,Research\,in\,Fundamental\,Science\,and\,Engineering,\,University\,of\,Tsukuba,\,Tsukuba,\,Ibaraki\,305,\,Japan}
\author{Xiaofeng Luo}
\affiliation{Key\,Laboratory\,of\,Quark\,and\,Lepton\,Physics\,(MOE)\,and\,Institute\,of\,Particle\,Physics,\,Central\,China\,Normal\,University,\,Wuhan\,430079,\,China}

%%%%%%%%%%%%%%%%%%%%%%%%%%%%%%%%%%%%%%%%%%%%%%%%%%%%%%%%%%%%%%%%%%%%%%%%%%%%%%%%%%%%%%%%%%%%%%%%%%%%%%%%%%%%%%%%%%
\begin{abstract}
We show the importance of separated efficiency corrections between positively and negatively charged particles 
for cumulant calculations by Monte Carlo toy models and analytical calculations.
Our results indicate that $S\sigma$ in published net-proton results from the STAR experiment will be suppressed about 5 to 10\% in central collisions,
 and 10 to 20\% in peripheral collisions at the beam energy of $\sqrt{s_{NN}}$~=~62.4 and 200~GeV if the separated efficiencies are used to efficiency correction.
%It is very important to use the separated efficiencies, not the averaged one.
\end{abstract}
%%%%%%%%%%%%%%%%%%%%%%%%%%%%%%%%%%%%%%%%%%%%%%%%%%%%%%%%%%%%%%%%%%%%%%%%%%%%%%%%%%%%%%%%%%%%%%%%%%%%%%%%%%%%%%%%%%

\pacs{Valid PACS appear here}% PACS, the Physics and Astronomy
                             % Classification Scheme.
%\keywords{Suggested keywords}%Use showkeys class option if keyword
                              %display desired
\maketitle
%%%%%%%%%%%%%%%%%%%%%%%%%%%%%%%%%%%%%%%%%%%%%%%%%%%%%%%%%%%%%%%%%%%%%%%%%%%%%%%%%%%%%%%%%%%%%%%%%%%%%%%%%%%%%%%%%%
\section{Introduction}
%%%%%%%%%%%%%%%%%%%%%%%%%%%%%%%%%%%%%%%%%%%%%%%%%%%%%%%%%%%%%%%%%%%%%%%%%
\subsection{Motivation}
Cumulants of conserved quantities~(net-charge, net-baryon or net-strangeness) are a powerful tool for searching the QCD critical point. 
Theoretically, cumulants are proportional to the power of correlation length 
and directly connected to the susceptibilities~\cite{fluctuation,correlation,susceptibility}. 
At the STAR experiment, cumulant ratios of net-proton and net-charge multiplicity distributions have been measured as a function of beam energy~\cite{net_proton,net_charge}. 
The results of net-proton cumulants and their ratios suggest that there might be something interesting around $\sqrt{s_{NN}}$~=~20~GeV, 
but more statistics is still necessary at low beam energy region due to their large statistical errors~\cite{net_proton}. The results of net-charge are consistent with statistical baselines~\cite{net_charge}.
Charged particles are measured by the Time Projection Chamber~(TPC). Tracking efficiency of the TPC for protons is about 70--90\% at midrapidity ($|y|~<~0.5$) in the range $0.4~<~p_{T}~<~0.8~$GeV/$c$, which depends on centralities and beam energies. In order to take into account the effect of finite tracking efficiency on net-proton cumulants, efficiency corrections~\cite{eff_koch,eff_kitazawa} are utilized, 
which assume the binomial response of tracking efficiency. Efficiency correction formulas assuming the identical efficiency between positively and negatively charge particles shown in \cite{eff_koch} was used in \cite{net_proton,net_charge}.  
However, there is about a few percent difference of tracking efficiency between positively and negatively charged particles. 
The main goal of this paper is to study how the published results will be changed if the separated efficiencies are used, which  
will be shown by Monte Carlo toy models and analytical calculations. 
%%%%%%%%%%%%%%%%%%%%%%%%%%%%%%%%%%%%%%%%%%%%%%%%%%%%%%%%%%%%%%%%%%%%%%%%%
\subsection{Cumulants and their baselines}
At the STAR experiment, cumulants up to fourth order were measured, and cumulant ratios~(which can be also written in terms of moments) defined as below were measured as a function of beam energies, 
\begin{eqnarray}
	\it{S}\sigma&=&\frac{C_{3}}{C_{2}}=\frac{\chi_{3}}{\chi_{2}}, \\
	\kappa\sigma^{2}&=&\frac{C_{4}}{C_{2}}=\frac{\chi_{4}}{\chi_{2}},
\end{eqnarray}
where $\sigma$, $\it{S}$ and $\kappa$ are the second to fourth order moments respectively, and $\chi_{n}$ is the $n$-th order susceptibility~\cite{susceptibility}. 
As cumulants are extensive variables, the volume effects can be canceled by taking their ratios.  
Theoretically, these variables are predicted to diverge near around the QCD critical point~\cite{correlation}. 
Experimentally, these observables are compared with statistical baseline which is called Skellam distribution in order to find a non-monotonic signal~\cite{nonmonotonic}. 
Skellam distribution is the difference between two independent Poisson distributions. 
Odd and even order cumulants of Skellam distribution are expressed in terms of mean parameter of Poisson distribution.
\begin{eqnarray}
	C_{\rm odd}&=&\mu_{+}-\mu_{-},
	\label{eq:skellam_odd} \\
	C_{\rm even}&=&\mu_{+}+\mu_{-},
	\label{eq:skellam_even}
\end{eqnarray}
where $\mu_{\pm}$ is the mean parameter of Poisson distribution for positively and negatively charged particles. 
Then the statistical baselines of cumulant ratios can be expressed as
\begin{eqnarray}
	&&\frac{C_{3}}{C_{2}}\biggl|_{Skellam}=\frac{\mu_{+}-\mu_{-}}{\mu_{+}+\mu_{-}},
	\label{eq:skellam_c32} \\
	&&\frac{C_{4}}{C_{2}}\biggl|_{Skellam}=1.
	\label{eq:skellam_c42}
\end{eqnarray}
Note that the baseline of $C_{3}/C_{2}$ can change with centrality and beam energy, 
while the baseline of $C_{4}/C_{2}$ is unity by definition.

\section{Analysis}
In this section, we discuss the potential problems of averaged efficiency correction by comparison with separated efficiency corrections. 
First, we show explicit expressions of cumulants for averaged and separated efficiency corrections in \ref{subsec2_A}.
Second, we demonstrate simple Monte Carlo toy models assuming Skellam distributions in \ref{subsec2_B}. 
Finally, we explain the results from \ref{subsec2_B} using analytical calculations in \ref{subsec2_C}.
%%%%%%%%%%%%%%%%%%%%%%%%%%%%%%%%%%%%%%%%%%%%%%%%
\subsection{Efficiency correction\label{subsec2_A}}
In order to discuss the analytical formulas of efficiency correction, we use the recursive expressions of moments and cumulants,
\begin{eqnarray}
\mu_{n}&=&\Bigl<\bigl(M_{+}-M_{-}\bigr)^{n}\Bigr>=\sum_{r=0}^{n}\binom{n}{r}(-1)^{r}\Bigl<M_{+}^{n-r}M_{-}^{r}\Bigr>, \nonumber \\
	\label{eq:noncentral} \\
C_{n}&=&\mu_{n}-\sum_{r=1}^{n-1}\binom{n-1}{r-1}C_{r}\mu_{n-r}, 
	\label{eq:cumulant}\\
\binom{n}{r}&=&\frac{n!}{r!(n-r)!},
\end{eqnarray}
where $\mu_{n}$ is the $n$-th order non-central moment, $C_{n}$ denotes the $n$-th order cumulant, $M_{\pm}$ is the number of observed particles
 and brackets represent the average over many events. 
Once non-central moments up to $n$-th order are calculated by Eq.~(\ref{eq:noncentral}), one can immediately calculate the $n$-th order cumulant 
recursively by Eq.~(\ref{eq:cumulant}). 
%Therefore, we don't need to write down the explicit formulas for efficiency correction because these recursive formulas make it very easy to calculate cumulants using the programming language. In this section we will show the explicit formulas up to third order to check the consistency with \cite{eff_koch}. See Appendix for the fourth order cumulant.
\par
Now let us define $N_{\pm}$ as the number of produced particles, $\varepsilon_{\pm}$ as the efficiency for positively and negatively charged particles.
$K_{n,{\rm ave}}$ and $K_{n,{\rm sep}}$ are the $n$-th order cumulant corrected by using averaged efficiency or separated efficiencies, respectively.
The first order cumulant can be simply corrected,
\begin{eqnarray}
	K_{1,{\rm sep}}&=&\bigl<N_{+}\bigr>-\bigl<N_{-}\bigr>=\frac{\bigl<M_{+}\bigr>}{\varepsilon_{+}} - \frac{\bigl<M_{-}\bigr>}{\varepsilon_{-}},
		       \label{eq:1stsep} \\
	K_{1,{\rm ave}}&=&\frac{\bigl<M_{+}\bigr>-\bigl<M_{-}\bigr>}{\varepsilon}=\frac{C_{1}}{\varepsilon},
		       \label{eq:1stave} \\
	\varepsilon&=&\frac{\varepsilon_{+}+\varepsilon_{-}}{2},
\end{eqnarray}
Equation~(\ref{eq:1stave}) can be easily obtained from Eq.~(\ref{eq:1stsep}) by replacing $\varepsilon_{\pm}$ to $\varepsilon$.
Similarly, the second order cumulant can be written as 
\begin{widetext}
\begin{eqnarray}
	K_{2,{\rm sep}}&=&\mu_{2}-\mu_{1}^{2}, \nonumber \\
		       &=&\bigl<N_{+}^{2}\bigr>-2\bigl<N_{+}N_{-}\bigr>+\bigl<N_{-}^{2}\bigr>-\Bigl(\bigl<N_{+}\bigr>-\bigl<N_{-}\bigr>\Bigr)^{2}, \nonumber \\
		       &=&\frac{\bigl<M_{+}^{2}\bigr>}{\varepsilon_{+}^{2}}-\frac{\bigl<M_{+}\bigr>}{\varepsilon_{+}^{2}}+\frac{\bigl<M_{+}\bigr>}{\varepsilon_{+}}-2\frac{\bigl<M_{+}M_{-}\bigr>}{\varepsilon_{+}\varepsilon_{-}}+\frac{\bigl<M_{-}^{2}\bigr>}{\varepsilon_{-}^{2}}-\frac{\bigl<M_{-}\bigr>}{\varepsilon_{-}^{2}}+\frac{\bigl<M_{-}\bigr>}{\varepsilon_{-}}-\biggl(\frac{\bigl<M_{+}\bigr>}{\varepsilon_{+}}-\frac{\bigl<M_{-}\bigr>}{\varepsilon_{-}}\biggr)^{2}. \label{eq:2ndsep} \\ 
	K_{2,{\rm ave}}&=&\frac{1}{\varepsilon^{2}}\biggl[\bigl<M_{+}^{2}\bigr>+\bigl<M_{-}^{2}\bigr>-2\bigl<M_{+}M_{-}\bigr>-\Big(\bigl<M_{+}\bigr> 
	-\bigl<M_{-}\bigr>\Bigr)^{2}\biggr]-\frac{1}{\varepsilon^{2}}\Bigl(\bigl<M_{+}\bigr>+\bigl<M_{-}\bigr>\Bigr)\Bigl(1-\varepsilon\Bigr), \nonumber \\ 
	&=&\frac{1}{\varepsilon^{2}}\Bigl[C_{2}-n\bigl(1-\varepsilon\bigr)\Bigr],
	\label{eq:2ndave}
\end{eqnarray}
\end{widetext}
where $n=f_{10}+f_{01}$ and $f_{ij}$ is the factorial moment~(see~\ref{app_A}).
%From the second to third line, we used the relationship between measured factorial moments and true ones as shown in Eq.~(\ref{eq:factorial_moment}). 
From the second to third line, we used Eq.~(\ref{eq:factorial_recursion}) shown in Appendix~\ref{app_A}. 
Equation~(\ref{eq:2ndave}) is the same as Eq.~(17) shown in \cite{eff_koch}. 
See Appendix \ref{app_DK3} for the third order cumulant. 
\subsection{Monte Carlo toy model\label{subsec2_B}}
\begin{figure}
\begin{center}
\includegraphics[width=85mm]{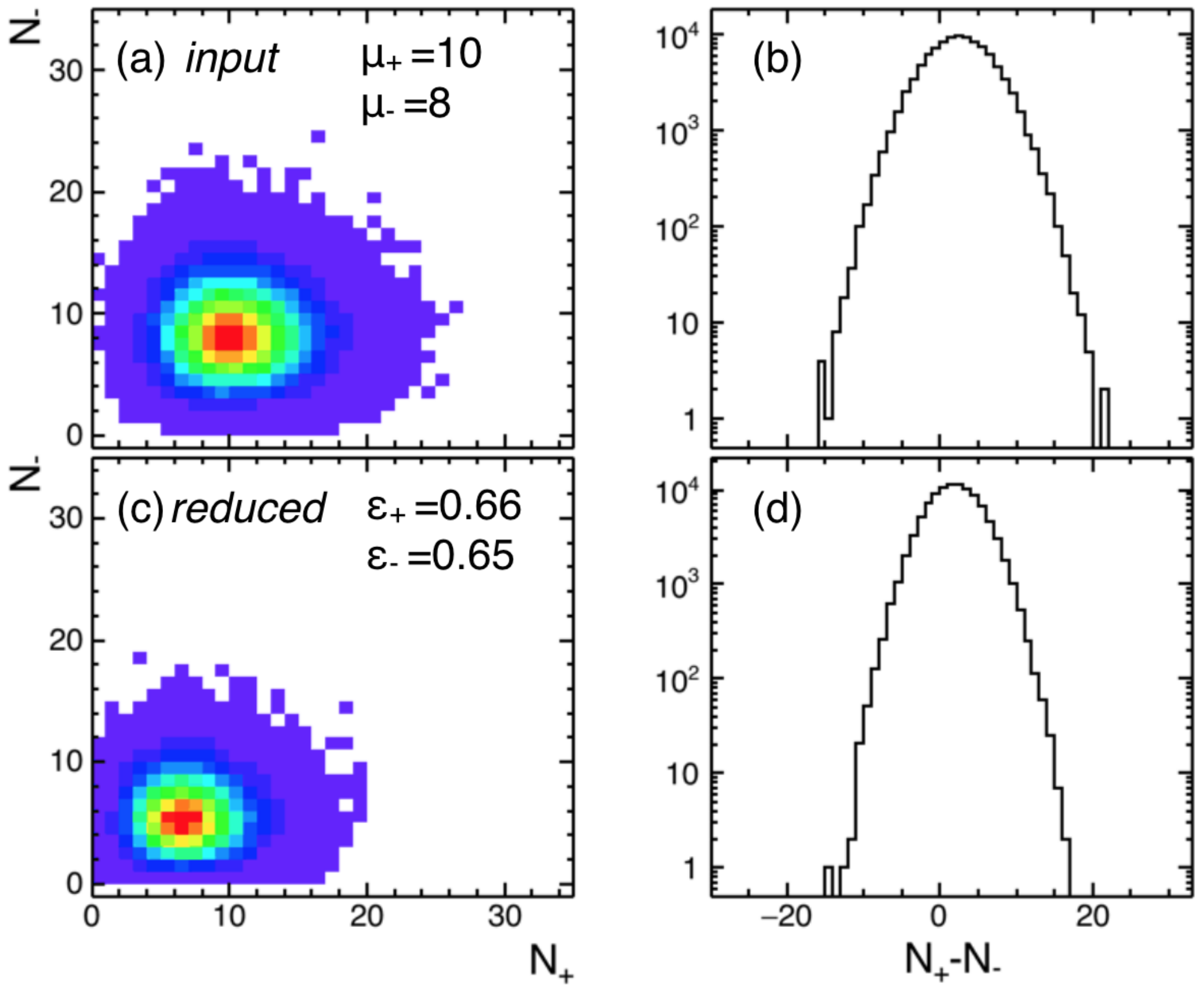}
\end{center}
\caption{(a)~Correlation between positively and negatively charged particles. (b)~Net-charge distribution which is calculated from (a). (c)~Correlation between positively and negatively charged particles, which are randomly sampled from (a) according to binomial efficiency. (d)~Net-charge distribution which is calculated from (c).} 
%This simulation was performed by generating 100K events and by 20 independent trials to evaluate the statistical uncertainties.}
\label{fig:simple}
\end{figure}

%%%%%%%%%%%%%%%%%%%%%%%%%%%%%%%%%%%%%%%%%%%%%%
Let us suppose two independent Poisson distributions, one is for positively charged particles and the other is for negatively charged particles, 
which are randomly generated according to parameters, $\mu_{+}$~=~10 and $\mu_{-}$~=~8, as shown in Fig.~\ref{fig:simple} (a) and (b). These particles are randomly sampled by each binomial efficiency, $\varepsilon_{+}$~=~0.66 and $\varepsilon_{-}$~=~0.65 as shown in Fig.~\ref{fig:simple} (c) and (d). Then, efficiency correction is applied by using averaged efficiency $\varepsilon=(\varepsilon_{+}+\varepsilon_{-})/2$ or separated efficiencies.
%%%%%%%%%%%%%%%%%%%%
%%%%%%%%%%%%%%%%%%%%
Efficiency corrected cumulants become
\begin{eqnarray}
K_{1,\rm ave}=2.14\pm0.02,~K_{1,\rm sep}=2.00\pm0.02 \label{eq:simple_result}
\end{eqnarray}
where the simulation was performed by generating 100K events, and by using 30 independent trials to evaluate statistical uncertainties on cumulants. 
We can see that the first order cumulant is increased 7\% from Skellam baseline~($K_{1}$~=~2) if we use averaged efficiency. 
Of course, the separated efficiency correction gives a consistent result with Skellam baseline. 
It is surprising that about 1\% difference of efficiencies leads to 7\% deviation of $C_{1}$ from input value.
%%%%%%%%%%%%%%%%%%%%%%%%%%%%%%%%%%%%%%%%%%%%%%
\par
\begin{figure}
\begin{center}
\includegraphics[width=85mm]{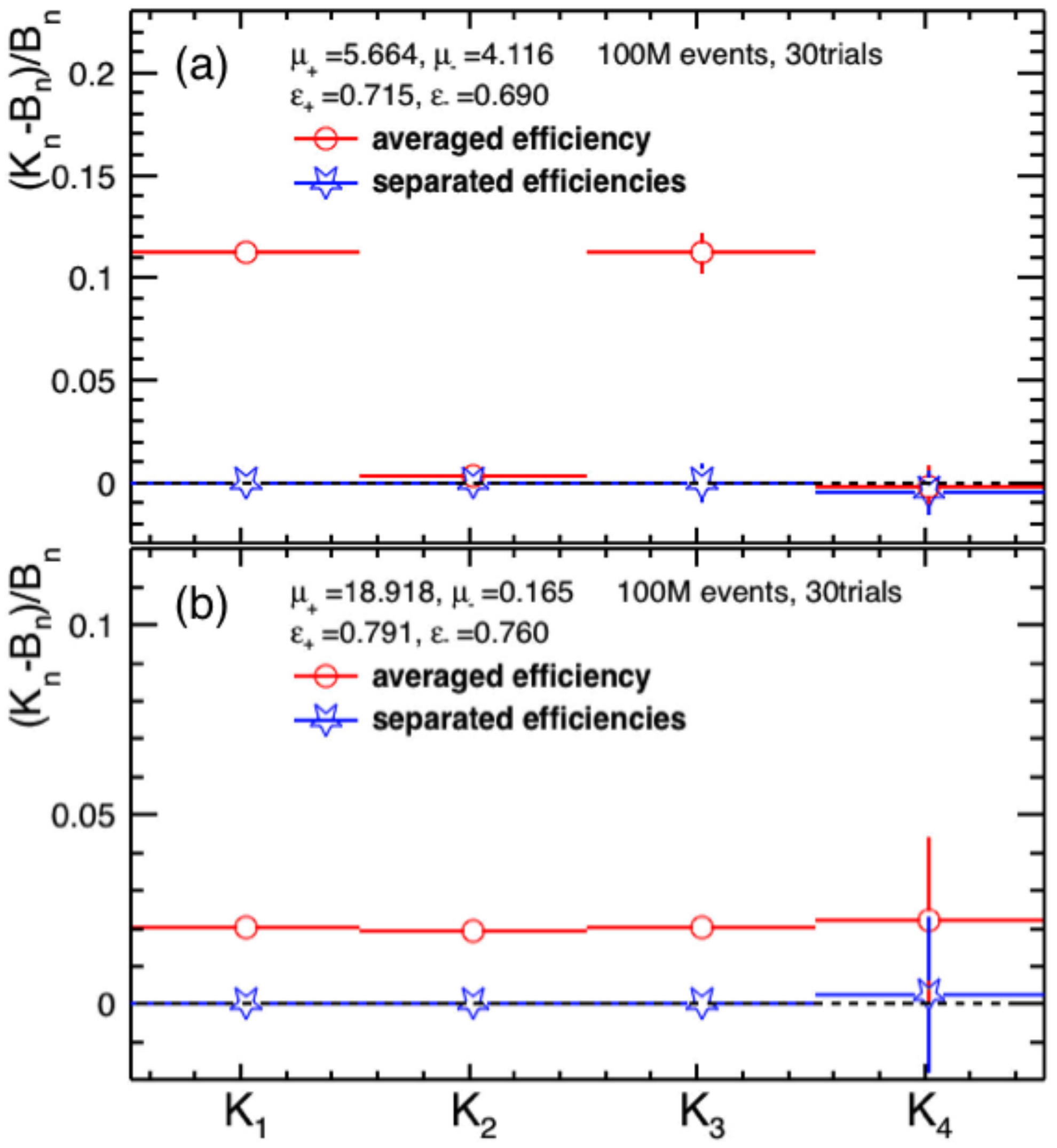}
\end{center}
\caption{(color online) Deviation of cumulants from input value assuming (a)~$\sqrt{s_{NN}}$~=~200~GeV and (b)~$\sqrt{s_{NN}}$~=~7.7~GeV, where $\mu_{\pm}$ and $\varepsilon_{\pm}$ were taken from \cite{net_proton,eff_proceedings}.
 $K_{n}$ represents the $n$-th order corrected cumulant and $B_{n}$ represents the Skellam baseline of the $n$-th order cumulant.}
\label{fig:AOS}
\end{figure}

We focus on the relative deviation of corrected cumulants from the Skellam baselines, $(K_{n}-B_{n})/B_{n}$, 
where $K_{n}$ represents the $n$-th order corrected cumulant and $B_{n}$ represents the Skellam baseline which is determined only by Poisson parameter as shown in Eqs.~(\ref{eq:skellam_odd}) and (\ref{eq:skellam_even}). 
This value should be zero if the efficiency correction works well.
Figure~\ref{fig:AOS} shows the relative deviation as a function of each order of cumulants and cumulant ratios, where the parameter of Poisson distribution $\mu_{+}$~($\mu_{-}$) is the efficiency corrected first order cumulants of proton~(antiproton) multiplicity distribution at (a)~$\sqrt{s_{NN}}$~=~200~GeV and (b)~$\sqrt{s_{NN}}$~=~7.7~GeV taken from \cite{net_proton}. 
Efficiency $\varepsilon_{\pm}$ was taken from \cite{eff_proceedings}. The simulation was performed by generating 100M events with 30 independent trials.
The results of separated efficiency corrections are consistent with zero for all the order cumulants and cumulant ratios, which confirms the validity of this efficiency correction. 
In case of averaged efficiency at $\sqrt{s_{NN}}$~=~200~GeV~(see Fig.~\ref{fig:AOS} (a)), however, $K_{1}$ and $K_{3}$ systematically increase about 10\% from input value, 
while there is very small deviation for $K_{2}$ and $K_{4}$. 
By contrast at $\sqrt{s_{NN}}$~=~7.7~GeV~(see Fig.~\ref{fig:AOS}~(b)), $K_{1}$ to $K_{4}$ increase about 2\%.
We will verify this observation by analytical calculations at next section.
%%%%%%%%%%%%%%%%%%%%
%%%%%%%%%%%%%%%%%%%%%%%%%%%%%%%%%%%%%%%%%%%%%%%%
\subsection{Analytical calculation\label{subsec2_C}}
Now we show how the results of averaged efficiency deviates from input value for the first and second order cumulants. 
First, we rewrite Eq.~(\ref{eq:1stsep}) as below:
\begin{eqnarray} 
	K_{1,\rm sep}=\frac{\bigl<M_{+}\bigr>}{\varepsilon+\Delta\varepsilon}-\frac{\bigl<M_{-}\bigr>}{\varepsilon-\Delta\varepsilon},
\end{eqnarray} 
where
\begin{eqnarray} 
	\varepsilon=\frac{\varepsilon_{+}+\varepsilon_{-}}{2},~\Delta\varepsilon=\frac{\varepsilon_{+}-\varepsilon_{-}}{2}. \nonumber
\end{eqnarray} 
We can apply Taylor expansion around $\Delta\varepsilon=0$ to Eq.~(\ref{eq:1stsep}) under the assumption that $\Delta\varepsilon$ is negligible, we obtain 
\begin{eqnarray}
	K_{1, \rm sep}(\Delta\varepsilon)&\approx&K_{1,\rm sep}(0) 
	+ \frac{\partial K_{1,\rm sep}}{\partial \Delta\varepsilon}\Bigl|_{\Delta\varepsilon=0}\Delta\varepsilon + \mathcal{O}(\Delta\varepsilon^{2}) \nonumber \\
	&\approx&\frac{\bigl<M_{+}\bigr>-\bigl<M_{-}\bigr>}{\varepsilon} - \frac{\bigl<M_{+}\bigr>+\bigl<M_{-}\bigr>}{\varepsilon^{2}}\Delta\varepsilon \nonumber \\
	&=&K_{1, \rm ave} - \frac{\bigl<M_{+}\bigr>+\bigl<M_{-}\bigr>}{\varepsilon^{2}}\Delta\varepsilon.
\end{eqnarray}
Thus,
\begin{eqnarray}
	\Delta K_{1}&=&K_{1,\rm ave}-K_{1,\rm sep} \nonumber \\
	      &\approx&\frac{\Delta\varepsilon}{\varepsilon^{2}}\Bigl(\bigl<M_{+}\bigr>+\bigl<M_{-}\bigr>\Bigr).
		\label{eq:delta_K1}
\end{eqnarray}
\begin{widetext}
Similarly for $\Delta K_{2}$,
\begin{eqnarray}
	\Delta K_{2}&=&K_{2,\rm ave}-K_{2,\rm sep} \nonumber \\
	      &\approx&-\frac{2\Delta\varepsilon}{\varepsilon^{2}}\Biggl[\frac{X_{+}-X_{-}}{\varepsilon}-\frac{1}{2}\Bigl(\bigl<M_{+}\bigr>-\bigl<M_{-}\bigr>\Bigr)\Biggr],
		\label{eq:delta_K2}
\end{eqnarray}
where ${X}_{\pm}=\bigl<M_{\pm}\bigr>-\bigl<M_{\pm}^{2}\bigr>+\bigl<M_{\pm}\bigr>^{2}$.  See Appendix \ref{app_DK3} for the analytical expression of $\Delta K_{3}$. 
From Eqs.~(\ref{eq:delta_K1}) and (\ref{eq:delta_K2}), we can intuitively predict that the deviation of odd order cumulants is proportional to the sum of function 
of multiplicity, 
and the deviation of even order cumulants is proportional to the difference of function of multiplicity.
%%%%%%%%%%%%%%%%%%%%%%%%%%%%%%%%%%%%%%%%%%%%%%%%%%%%%%%%%%%%%%%%%%%%%%%%%%%%%%%%%%%%
%\paragraph*{General expression}
\par
Since it is very cumbersome to calculate more than third order cumulant, we consider the general properties of difference between odd and even order cumulants. 
From the definitions of cumulants and non-central moments (see Eqs.~(\ref{eq:noncentral}) and (\ref{eq:cumulant})) and the analytical formula shown in Eqs.~(\ref{eq:1stsep}), (\ref{eq:2ndsep}) and (\ref{eq:3rdsep}), even order cumulants and non-central moments are $symmetrical$ with 
$M_{+}$ and $M_{-}$, while odd orders are $antisymmetrical$. Therefore, they can be generally expressed as
\begin{eqnarray}
C_{\rm 2m+1,\rm sep}&=&f_{\rm 2m+1}(M_{+},M_{-})-f_{\rm 2m+1}(M_{-},M_{+}), \\
C_{\rm 2m,\rm sep}&=&f_{\rm 2m}(M_{+},M_{-})+f_{\rm 2m}(M_{-},M_{+}), 
\end{eqnarray}
where $f_{\rm 2m+1}$ and $f_{\rm 2m}$ are functions which satisfies above two equations for odd and even order cumulants respectively. 
Similarly we define the efficiency corrected cumulants as
\begin{eqnarray}
 K_{\rm 2m+1,\rm sep}&&=F_{\rm 2m+1}[(M_{+},\varepsilon_+),(M_{-},\varepsilon_-)] - F_{\rm 2m+1}[(M_{-},\varepsilon_-),(M_{+},\varepsilon_+)],
	\label{eq:general_odd} \\ 
 K_{\rm 2m,\rm sep}&&=F_{\rm 2m}[(M_{+},\varepsilon_+),(M_{-},\varepsilon_-)] + F_{\rm 2m}[(M_{-},\varepsilon_-),(M_{+},\varepsilon_+)],  
	\label{eq:general_even}
\end{eqnarray}
where $F$ is the efficiency corrected result of $f$, which can be expressed in terms of $M_{\pm}$ and $\varepsilon_{\pm}$.  
Replacing $\varepsilon_{\pm}$ to $\varepsilon\pm\Delta\varepsilon$, we obtain 
\begin{eqnarray}
&&K_{\rm 2m+1,\rm sep}\approx F_{\rm 2m+1}[(M_{+},\varepsilon),(M_{-},\varepsilon)]-F_{\rm 2m+1}[(M_{-},\varepsilon),(M_{+},\varepsilon)] \nonumber \\
&&+\Biggl[\left.\frac{\partial{F_{\rm 2m+1}}[(M_{+},\varepsilon+\Delta\varepsilon),(M_{-},\varepsilon-\Delta\varepsilon)]}{\partial\Delta\varepsilon}\right|_{\Delta\varepsilon=0} - \left.\frac{\partial{F_{\rm 2m+1}}[(M_{-},\varepsilon-\Delta\varepsilon),(M_{+},\varepsilon+\Delta\varepsilon)]}{\partial\Delta\varepsilon}\right|_{\Delta\varepsilon=0}\Biggr]\Delta\varepsilon+\mathcal{O}(\Delta\varepsilon^{2}) \nonumber \\
%&\approx&K_{\rm odd,\rm ave}+\left.\frac{\partial{F}[(M_{+},\varepsilon+\Delta\varepsilon),(M_{-},\varepsilon-\Delta\varepsilon)]}{\partial\Delta\varepsilon}\right|_{\Delta\varepsilon=0}\Delta\varepsilon \nonumber \\
%&-& \left(\frac{\partial\Delta\varepsilon'}{\partial\Delta\varepsilon}\right)\left.\frac{\partial{F}[(M_{-},\varepsilon+\Delta\varepsilon'),(M_{+},\varepsilon-\Delta\varepsilon')]}{\partial\Delta\varepsilon'}\right|_{\Delta\varepsilon'=0}\Delta\varepsilon, \label{eq:sugi1}
&& \approx K_{\rm 2m+1,\rm ave} \nonumber \\
&&+\Biggl[\left.\frac{\partial{F_{\rm 2m+1}}[(M_{+},\varepsilon+\Delta\varepsilon),(M_{-},\varepsilon-\Delta\varepsilon)]}{\partial\Delta\varepsilon}\right|_{\Delta\varepsilon=0} - \left(\frac{\partial\Delta\varepsilon'}{\partial\Delta\varepsilon}\right)\left.\frac{\partial{F_{\rm 2m+1}}[(M_{-},\varepsilon+\Delta\varepsilon'),(M_{+},\varepsilon-\Delta\varepsilon')]}{\partial\Delta\varepsilon'}\right|_{\Delta\varepsilon'=0}\Biggr]\Delta\varepsilon, \label{eq:sugi1}
\end{eqnarray}
where 
\begin{eqnarray}
\Delta\varepsilon'=-\Delta\varepsilon.
\end{eqnarray}
Then, we define $G(x,y,\varepsilon,\Delta\varepsilon)$ as derivative of $F[(x,\varepsilon_{+}),(y,\varepsilon_{-})]$ with respect to $a$,
\begin{eqnarray}
G(x,y,\varepsilon,\Delta\varepsilon)&=&\left.\frac{\partial{F}[(x,\varepsilon+a),(y,\varepsilon-a)]}{\partial{a}}\right|_{a=0}\Delta\varepsilon. 
\end{eqnarray}
Therefore, 
\begin{eqnarray}
\Delta K_{\rm 2m+1}&=&K_{\rm 2m+1,\rm ave}-K_{\rm 2m+1,\rm sep}, \nonumber \\
                  &=&G_{\rm 2m+1}(M_{+},M_{-},\varepsilon,\Delta\varepsilon) + G_{\rm 2m+1}(M_{-},M_{+},\varepsilon,\Delta\varepsilon). \label{eq:ana_gene_odd}
\end{eqnarray}
Similarly for even order cumulants, 
\begin{eqnarray}
\Delta K_{\rm 2m}&=&K_{\rm 2m,\rm ave}-K_{\rm 2m,\rm sep}, \nonumber \\
                   &=&{G_{\rm 2m}(M_{+},M_{-},\varepsilon,\Delta\varepsilon)-G_{\rm 2m}(M_{-},M_{+},\varepsilon,\Delta\varepsilon}). \label{eq:ana_gene_even}
\end{eqnarray}
Eqs.~(\ref{eq:ana_gene_odd}) and (\ref{eq:ana_gene_even}) indicate that the deviation of odd order 
cumulants is represented as sum of $G(M_{+},M_{-},\varepsilon,\Delta\varepsilon)$ and $G(M_{-},M_{+},\varepsilon,\Delta\varepsilon)$, 
while the deviation of even order cumulants is represented as difference between them. Confirmation of these calculations is shown in Appendix.~\ref{app_GE}. 
Note that Eqs.~(\ref{eq:ana_gene_odd}) and (\ref{eq:ana_gene_even}) are valid for any probability distribution. 
\end{widetext}
%%%%%%%%%%%%%%%%%%%%%%%%%%%%%%%%%%%%%%%%%%%%%%%%%%%%%%%%%%%%%%%%%%%%%%%%%%%%%%%%%%%%
\par
Now let us assume that both positively and negatively charged particles follow Poisson distribution. From the definition of Skellam distribution shown in Eqs.~(\ref{eq:skellam_odd}) and (\ref{eq:skellam_even}), 
we can easily derive the following expressions:
\begin{eqnarray}
\Delta K_{\rm odd}&\approx&\frac{\Delta\varepsilon}{\varepsilon^{2}}\Bigl(\bigl<M_{+}\bigr>+\bigl<M_{-}\bigr>\Bigr), \label{eq:delta_odd}\\
\Delta K_{\rm even}&\approx&\frac{\Delta\varepsilon}{\varepsilon^{2}}\Bigl(\bigl<M_{+}\bigr>-\bigl<M_{-}\bigr>\Bigr). \label{eq:delta_even}
\end{eqnarray}
Equations~(\ref{eq:delta_odd}) and (\ref{eq:delta_even}) indicate that the deviation of odd order cumulant is proportional to the sum of multiplicity, while the deviation of even order cumulant is proportional to the difference of multiplicity.
Using parameters shown in Fig.~\ref{fig:AOS}~(a) we can estimate $\Delta K_{n}$ analytically. 
Comparison between toy model simulations and analytical calculations are summarized in Tab.~\ref{tab:delta_200} and \ref{tab:delta_7}. 
These results are roughly consistent, although $\Delta K_{4}$ has large statistical errors.
%%%%%%%%%%%%%%%%%%%
\begin{table}
\caption{\label{tab:delta_200}Comparison of $\Delta K_{n}$ between toy model simulations and analytical calculations assuming Skellam distribution at $\sqrt{s_{NN}}$~=~200~GeV.}
\begin{ruledtabular}
\begin{tabular}{ccc}
$n$-th order & MC toy model & analytical calculation \\ \hline 
1 & $0.1123\pm0.0002 $ & 0.112  \\ 
2 & $0.0028\pm0.0002 $ & 0.004  \\ 
3 & $0.112\pm0.010 $ & 0.112  \\ 
4 & $-0.002\pm0.010 $ & 0.004 \\ 
\end{tabular}
\end{ruledtabular}
\end{table}
%%%%%%%%%%%%%%%%%%%
\begin{table}
\caption{\label{tab:delta_7}Comparison of $\Delta K_{n}$ between toy model simulations and analytical calculations assuming Skellam distribution at $\sqrt{s_{NN}}$~=~7.7~GeV.}
\begin{ruledtabular}
\begin{tabular}{ccc}
$n$-th order & MC toy model & analytical calculation \\ \hline 
1 & $0.0202\pm2.4\times10^{-5} $ & 0.021  \\
2 & $0.0195\pm0.0002 $ & 0.020  \\ 
3 & $0.020\pm0.002 $ & 0.021  \\ 
4 & $0.022\pm0.022 $ & 0.020  \\ 
\end{tabular}
\end{ruledtabular}
\end{table}
%%%%%%%%%%%%%%%%%%%

\section{Results}
\begin{figure}
\begin{center}
\includegraphics[width=85mm]{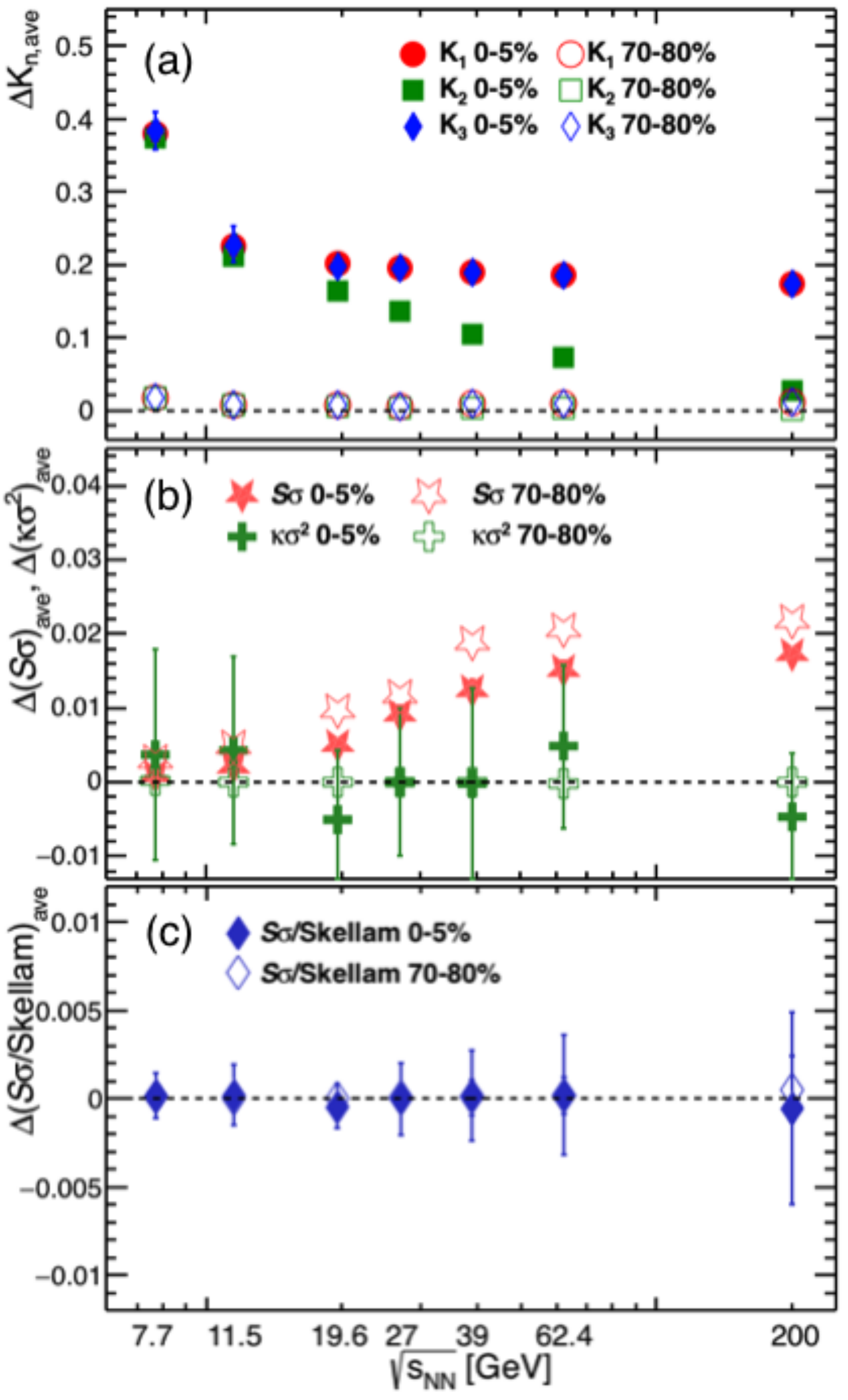}
\end{center}
\caption{(color online) Deviation of (a)~$K_{1}$, $K_{2}$ and $K_{3}$ (b)~$S\sigma$ and $\kappa\sigma^{2}$ (c)~$S\sigma/Skellam$ from input value as a function of beam energy. They are defined in Eqs.~(\ref{eq:deviation_ave})--(\ref{eq:deviation_ave_skewsigmaskellam}).  
Closed symbols represent the result assuming 0-5\% central collisions, and open symbols represent the result assuming 70-80\% peripheral collisions. Dashed lines represent the baseline for efficiency correction (=0). }
\label{fig:BES}
\end{figure}

As we discussed in previous section, the deviation of $K_{n,\rm ave}$ depends on the multiplicity, which indicates that it also depends on the beam energies. 
Fig.~\ref{fig:BES} shows the deviation
of (a)~$K_{1}$, $K_{2}$ and $K_{3}$ (b)~$S\sigma$ and $\kappa\sigma^{2}$ (c)~$S\sigma/Skellam$ from input value as a function of beam energy for central and peripheral collisions, where these deviations are defined as
\begin{eqnarray}
 \Delta K_{n,\rm ave}&=&K_{n,\rm ave}-B_{n}, \label{eq:deviation_ave} \\
 \Delta(S\sigma)_{\rm ave}&=&(S\sigma)_{ave}-B_{S\sigma}, \label{eq:deviation_ave_skewsigma} \\
 \Delta(\kappa\sigma^{2})_{\rm ave}&=&(\kappa\sigma^{2})_{ave}-B_{\kappa\sigma^{2}}, \label{eq:deviation_ave_kappasigma} \\
 \Delta(S\sigma/Skellam)_{\rm ave}&=&(S\sigma/Skellam)_{ave}-B_{S\sigma/Skellam}, \nonumber \\ \label{eq:deviation_ave_skewsigmaskellam}
\end{eqnarray}
where $K_{n,\rm ave}$, $(S\sigma)_{\rm ave}$, $(\kappa\sigma^{2})_{\rm ave}$ and $(S\sigma/Skellam)_{\rm ave}$ denote the $n$-th order cumulant, $S\sigma$, $\kappa\sigma^{2}$ and $S\sigma/Skellam$ corrected by using averaged efficiency. $B_{n}$, $B_{S\sigma}$, $B_{\kappa\sigma^{2}}$ and $B_{S\sigma/Skellam}$ are their Skellam baselines. 
From Fig.~\ref{fig:AOS} one can see that the separated efficiency correction gives a correct value which equals to the Skellam baseline. Thus, we used Skellam baselines in Eqs.~(\ref{eq:deviation_ave})--(\ref{eq:deviation_ave_skewsigmaskellam}) for simplicity.
As one can see in Fig.~\ref{fig:BES} (a), the deviation decreases as beam energy, and different behavior can be observed between odd and even order as discussed in previous section. 
At high beam energies, the deviation of $K_{2}$ from input value is close to $0$ while the deviations for $K_{1}$ and $K_{3}$ stays about $0.2$ because net-baryon is very small at midrapidity, $\overline{p}/p\simeq0.727$ at $\sqrt{s_{NN}}$~=~200~GeV~\cite{net_proton}.
At low beam energies, the deviation of $K_{2}$ is as large as that of $K_{1}$ and $K_{3}$ due to large net-baryon, $\overline{p}/p\simeq0.009$ at $\sqrt{s_{NN}}$~=~7.7~GeV~\cite{net_proton}. 
This difference between odd and even order leads to the behavior of $S\sigma$ and $\kappa\sigma^{2}$ shown in Fig.~\ref{fig:BES} (b). The deviation of $S\sigma$ increases as beam energy, and the deviation of $\kappa\sigma^{2}$ becomes almost zero, 
because $\kappa\sigma^{2}$ is the ratio of even order cumulants. As one can see in Fig.~\ref{fig:BES}~(c), the deviation of $S\sigma/Skellam$ is zero for all over the beam energy. This is because the Skellam baseline of $S\sigma$ is also affected by averaged efficiency~(see \ref{ssigmaskellam} for details). 
%%%%%%%%%%%%%%%%%%%
%%%%%%%%%%%%%%%%%%%%
\par
\begin{figure}
\begin{center}
\includegraphics[width=85mm]{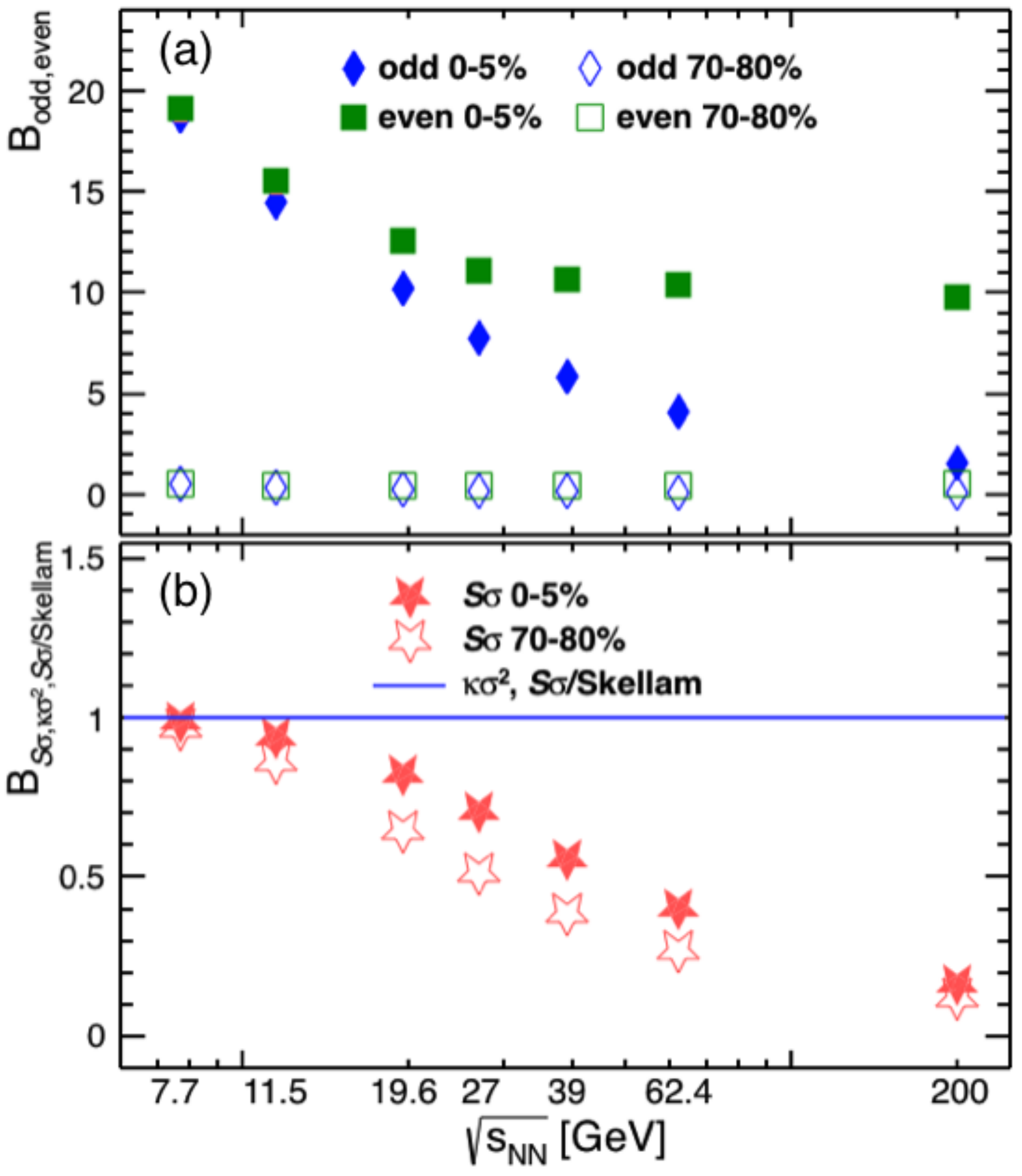}
\end{center}
\caption{(color online) Skellam baseline of (a)~odd and even order cumulants (b)~$S\sigma$, $\kappa\sigma^{2}$ and $S\sigma/Skellam$ as a function of beam energy. Closed symbols represent the result assuming 0-5\% central collisions, and open symbols represent the result assuming 70-80\% peripheral collisions. Baselines of $\kappa\sigma^{2}$ and $S\sigma/Skellam$ are represented as a solid blue line since they are unity by definition.} 
\label{fig:baselineBES}
\end{figure}

In order to discuss the relative deviation as a function of beam energy, we also studied the beam energy dependence of Skellam baseline, 
which can be directly calculated according to Eqs.~(\ref{eq:skellam_odd})--(\ref{eq:skellam_c42}) without MC toy models.
Figure~\ref{fig:baselineBES} shows Skellam baselines of (a)~odd and even order cumulant (b)~$S\sigma$, $\kappa\sigma^{2}$ and $S\sigma/Skellam$ as a function of beam energy for central and peripheral collisions. 
We note that the baselines of $\kappa\sigma^{2}$ and $S\sigma/Skellam$ are unity by their definitions. 
%%%%%%%%%%%%%%%%%%%%
%%%%%%%%%%%%%%%%%%%%
\par
\begin{figure}
\begin{center}
\includegraphics[width=85mm]{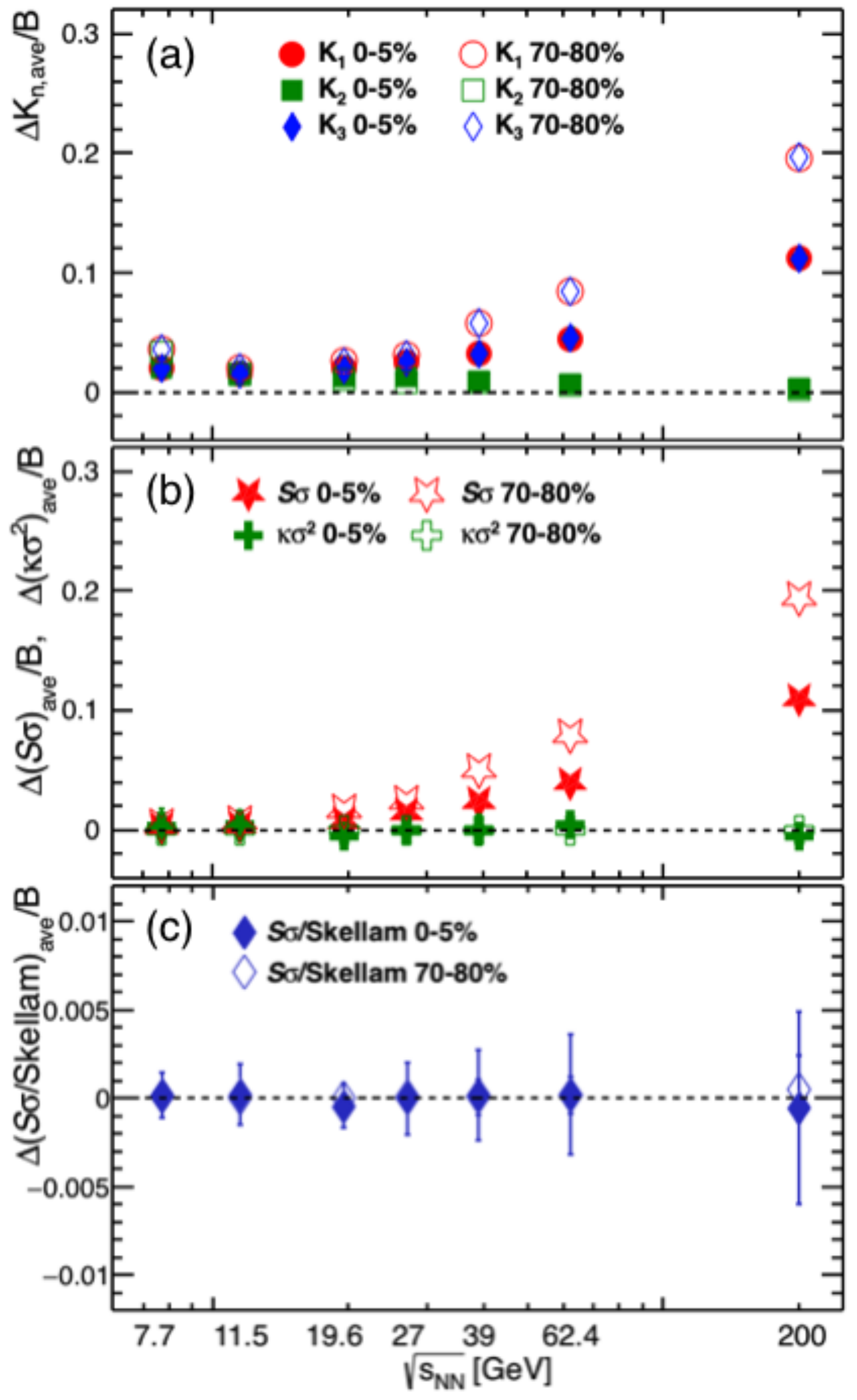}
\end{center}
\caption{(color online) Relative deviation of (a)~$K_{1}$, $K_{2}$ and $K_{3}$ (b)~$S\sigma$ and $\kappa\sigma^{2}$ (c)~$S\sigma/Skellam$ from input value as a function of beam energy, which corresponds to the ratio of Fig.~\ref{fig:BES} to Fig.~\ref{fig:baselineBES}. Closed symbols represent the result assuming 0-5\% central collisions, and open symbols represent the result assuming 70-80\% peripheral collisions. Dashed lines represent the baseline for efficiency correction (=0).} 
\label{fig:relative_bes}
\end{figure}

Figure~\ref{fig:relative_bes} shows the relative deviation of (a)~$K_{1}$, $K_{2}$ and $K_{3}$ (b)~$S\sigma$ and $\kappa\sigma^{2}$ (c)~$S\sigma/Skellam$ from input value as a function of beam energy, which corresponds to the ratio of Fig.~\ref{fig:BES} to Fig.~\ref{fig:baselineBES}. 
$K_{1}$ and $K_{3}$ deviate about 10 to 20\% in peripheral collisions and 5 to 10\% in central collisions at $\sqrt{s_{NN}}$~=~62.4 and 200~GeV, while
the deviation of $K_{2}$ is very small, which leads to the large deviation of $S\sigma$ and small deviation of $\kappa\sigma^{2}$ and $S\sigma/Skellam$. 
$S\sigma/Skellam$ doesn't deviate for all over the beam energy. 

\section{Discussion}
\subsection{$S\sigma/Skellam$\label{ssigmaskellam}}
As can be seen in Fig.~\ref{fig:AOS}, $S\sigma$ deviates from input value because it is defined as the ratio of third to second order cumulant, 
nevertheless $S\sigma/Skellam$ doesn't deviate~(see Figs.~\ref{fig:BES}~(c) and \ref{fig:relative_bes}~(c)). This is because Skellam term is also affected if we use the averaged efficiency.
From Eq.~(\ref{eq:skellam_c32}), we obtain 
\begin{eqnarray}
%	\frac{C_{3}}{C_{2}}\biggl|_{Skellam,uncorr}&=&\frac{\bigl<M_{+}\bigr>-\bigl<M_{-}\bigr>}{\bigl<M_{+}\bigr>+\bigl<M_{-}\bigr>}, \\
%	\frac{C_{3}}{C_{2}}\biggl|_{Skellam,corr,\rm sep}&=&
%		\frac{\displaystyle{\frac{\bigl<M_{+}\bigr>}{\varepsilon_{+}}-\frac{\bigl<M_{-}\bigr>}{\varepsilon_{-}}}}{\displaystyle{\frac{\bigl<M_{+}\bigr>}{\varepsilon_{+}}+\frac{\bigl<M_{-}\bigr>}{\varepsilon_{-}}}}, \nonumber \\
					%	&=&\frac{\varepsilon_{-}\bigl<M_{+}\bigr>-\varepsilon_{+}\bigl<M_{-}\bigr>}{\varepsilon_{-}\bigl<M_{+}\bigr>+\varepsilon_{+}\bigl<M_{-}\bigr>}.
	\frac{C_{3}}{C_{2}}\biggl|_{Skellam}&=&\frac{\bigl<M_{+}\bigr>-\bigl<M_{-}\bigr>}{\bigl<M_{+}\bigr>+\bigl<M_{-}\bigr>}, \\
	\frac{K_{3}}{K_{2}}\biggl|_{Skellam,\rm sep}&=&\frac{\bigl<N_{+}\bigr>-\bigl<N_{-}\bigr>}{\bigl<N_{+}\bigr>+\bigl<N_{-}\bigr>}
						    =\frac{\displaystyle{\frac{\bigl<M_{+}\bigr>}{\varepsilon_{+}}-\frac{\bigl<M_{-}\bigr>}{\varepsilon_{-}}}}{\displaystyle{\frac{\bigl<M_{+}\bigr>}{\varepsilon_{+}}+\frac{\bigl<M_{-}\bigr>}{\varepsilon_{-}}}} \nonumber \\
						&=&\frac{\varepsilon_{-}\bigl<M_{+}\bigr>-\varepsilon_{+}\bigl<M_{-}\bigr>}{\varepsilon_{-}\bigl<M_{+}\bigr>+\varepsilon_{+}\bigl<M_{-}\bigr>}.
\end{eqnarray}
In case of averaged efficiency, we obtain
\begin{eqnarray}
	\frac{K_{3}}{K_{2}}\biggl|_{Skellam,\rm ave}&=&\frac{\displaystyle{\frac{\bigl<M_{+}\bigr>-\bigl<M_{-}\bigr>}{\varepsilon}}}{\displaystyle{\frac{\bigl<M_{+}\bigr>+\bigl<M_{-}\bigr>}{\varepsilon}}}, \nonumber \\
							&=&\frac{\bigl<M_{+}\bigr>-\bigl<M_{-}\bigr>}{\bigl<M_{+}\bigr>+\bigl<M_{-}\bigr>}=\frac{C_{3}}{C_{2}}\biggl|_{Skellam}, \nonumber \\
\end{eqnarray}
which means the Skellam term is still uncorrected if we use the averaged efficiency. Thus, the reason why $S\sigma/Skellam$ doesn't deviate from input value is because numerator~($S\sigma$) and denominator~(Skellam) are affected simultaneously.

\subsection{Weighted averaged efficiency}
\begin{figure}
\begin{center}
\includegraphics[width=85mm]{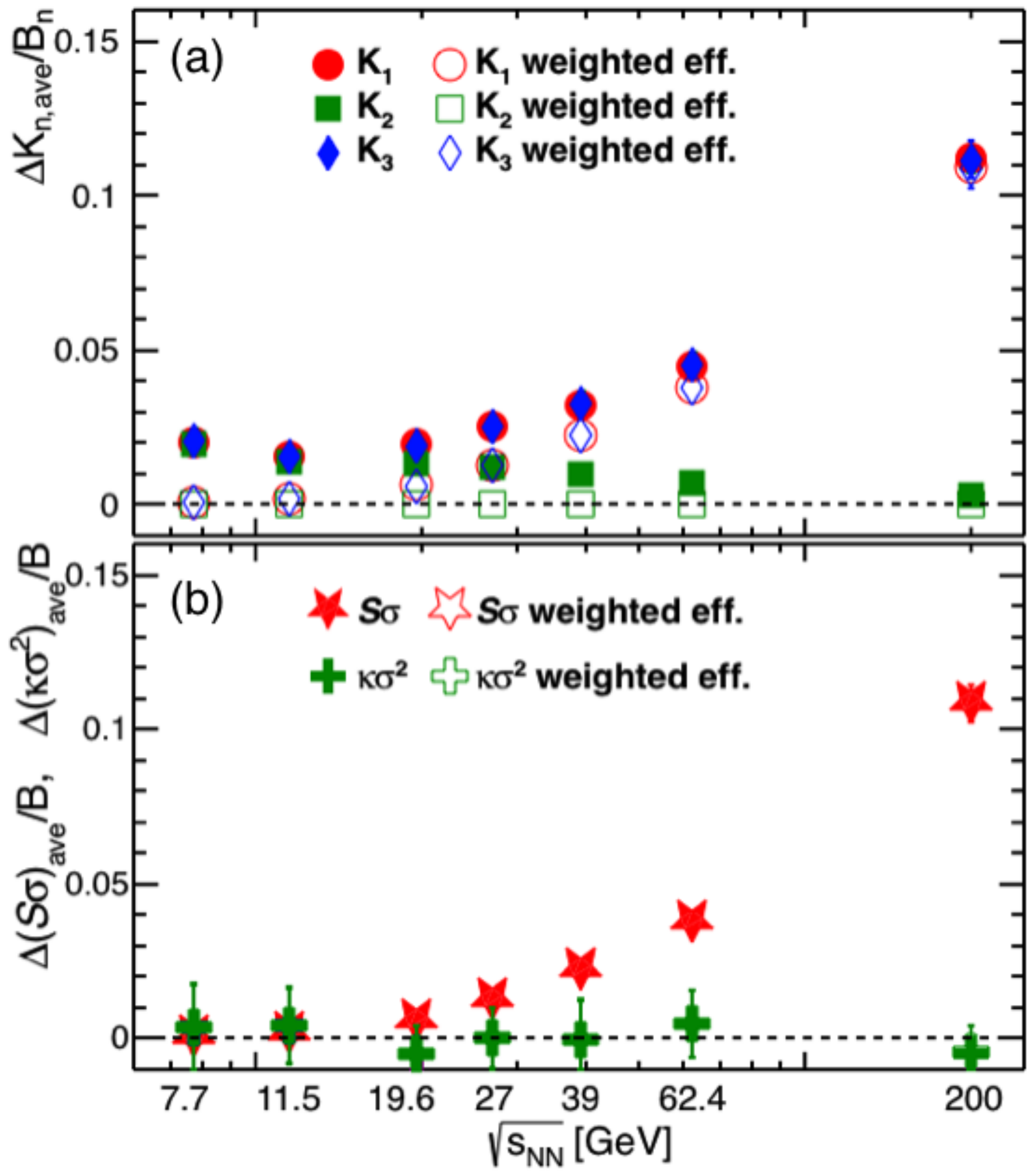}
\end{center}
\caption{(color online) Relative deviation of results corrected by using averaged efficiency and weighted averaged efficiency as a function of beam energy. Each panel shows (a)~$K_{1}$, $K_{2}$ and $K_{3}$ (b)~$S\sigma$ and $\kappa\sigma^{2}$. For $S\sigma$ and $\kappa\sigma^{2}$ the results of weighted averaged efficiency are overlapped with the results of averaged efficiency. Dashed lines represent the baseline for efficiency correction (=0).} 
\label{fig:weighted}
\end{figure}

Intuitively, it is more appropriate for averaged efficiency to be weighted by number of particles as
\begin{eqnarray}
	\varepsilon_{w}=\frac{\bigl<N_{+}\bigr>\varepsilon_{+}+\bigl<N_{-}\bigr>\varepsilon_{-}}{\bigl<N_{+}\bigr>+\bigl<N_{-}\bigr>},
\end{eqnarray}
where $N_{\pm}$ is the number of particles.
This {\it weighted averaged} efficiency should consider the effect of small number of anti-protons at low beam energy region.
MC toy models using $\varepsilon_{w}$ was also studied. Fig.~\ref{fig:weighted} shows the relative deviation from input value in two cases, one is calculated by averaged efficiency, the other is calculated by weighted averaged efficiency as a function of beam energy assuming 0-5\% central collisions. 
Weighted averaged efficiency gives better results than averaged efficiency for $K_{1}$, $K_{2}$ and $K_{3}$ at low beam energy region due to small number of anti-protons, while the results of weighted averaged efficiency deviate as large as the averaged efficiency at high beam energy region.
For $S\sigma$ the results of weighted averaged efficiency deviate as large as the averaged efficiency at all over the beam energy.
These results indicate that we should not use the averaged efficiency as well as weighted one.

\section{Summary}
Importance of separated efficiencies between positively and negatively charged particles was shown by Monte Carlo toy models and analytical calculations.
MC toy models assuming parameters from \cite{net_proton} indicate that, odd order cumulants and $S\sigma$ systematically deviate from input value about 10\% in case of averaged efficiency at $\sqrt{s_{NN}}$~=~200~GeV, while the deviation of even order cumulants is as large as odd order cumulants at $\sqrt{s_{NN}}$~=~7.7~GeV.
In order to understand the nature of this behavior, analytical calculation was also performed. Results indicate that the deviation of odd order cumulants is proportional to the sum of multiplicity, while the deviation of even order cumulants is proportional to the difference of multiplicity.
Moreover, beam energy dependence was studied assuming the published net-proton results. $S\sigma/Skellam$ and $\kappa\sigma^{2}$ don't deviate from input value for beam energies in which we studied. $S\sigma$ is enhanced about 5 to 10\% in central collisions and 10 to 20\% in peripheral collisions at $\sqrt{s_{NN}}$~=~62.4 and 200~GeV, and there is less than 5\% enhancement at $\sqrt{s_{NN}}$~=~7.7--39~GeV, where the QCD critical point is predicted to exist. Therefore, these enhancement arising from using the averaged efficiency don't change the conclusions in \cite{net_proton}. However, it is definitely the right way to use separated efficiencies, not the averaged one.
\par
In order to develop correction methods which assumes more realistic situation of heavy ion colliding experiments, 
more detailed characteristics were studied for efficiency correction, i.e. correction formulas imposing phase space dependent efficiencies~\cite{eff_psd_volker,eff_psd_kitazawa}, and the breaking of the binomial assumption of tracking efficiencies~\cite{binomial_breaking}. 
Recently at the STAR experiment, the $p_{T}$ region are extended up to $2.0$~GeV/c by including Time Of Flight for particle identification~\cite{eff_proceedings}, 
which leads to the low tracking efficiency at high $p_{T}$ region. This effect is corrected by dividing phase space of (anti)proton into high and low $p_{T}$ regions using the efficiency correction from \cite{eff_xiaofeng}.

\section*{Acknowledgment}
We thank Nu Xu and Masakiyo Kitazawa for active discussions with us, and the staff of the Collider-Accelerator and Physics Departments at BNL and STAR collaboration for their vital contributions. 
We also acknowledge support from MEXT and JSPS KAKENHI Grant Number 
25105504 and Super Global University Program in University of Tsukuba. 
X. Luo was supported in part by the MoST of China 973-Project No.~2015CB856901 and NSFC under grant No.~11575069.

\begin{widetext}
\section{Appendixes}
%%%%%%%%%%%%%%%%%%%%%%%%%%%%%%%%%%%%%%%%%%%%%%%%%%%%%%%%%%%%%%%%%%%%%
\subsection{Factorial moments\label{app_A}}
Factorial moments have a useful characteristics as shown below,
\begin{eqnarray}
	f_{ab}&=&\varepsilon_{+}^{a}\varepsilon_{-}^{b}F_{ab}, \label{eq:factorial_moment}\\
	f_{ab}&=&\sum_{M_{+}=a}^{\infty}\sum_{M_{-}=b}^{\infty} p(M_{+},M_{-})\frac{M_{+}!}{(M_{+}-a)!}\frac{M_{-}!}{(M_{-}-b)!}, \nonumber \\ \\
	F_{ab}&=&\sum_{N_{+}=a}^{\infty}\sum_{N_{-}=b}^{\infty} P(N_{+},N_{-})\frac{N_{+}!}{(N_{+}-a)!}\frac{N_{-}!}{(N_{-}-b)!}. \nonumber \\ 
\end{eqnarray}
%\if0
They can be also written in terms of Stirling number of the first kind.
\begin{eqnarray}
	f_{ab}&=&\sum_{i=0}^{a}\sum_{j=0}^{b} s(a,i)s(b,j)\bigl<M_{+}^{i}M_{-}^{j}\bigr>, \\
	F_{ab}&=&\sum_{i=0}^{a}\sum_{j=0}^{b} s(a,i)s(b,j)\bigl<N_{+}^{i}N_{-}^{j}\bigr>, 
\end{eqnarray}
thus,
\begin{eqnarray}
	\sum_{i=0}^{a}\sum_{j=0}^{b} s(a,i)s(b,j)\bigl<M_{+}^{i}M_{-}^{j}\bigr>=\varepsilon_{+}^{a}\varepsilon_{-}^{b}\sum_{i=0}^{a}\sum_{j=0}^{b} s(a,i)s(b,j)\bigl<N_{+}^{i}N_{-}^{j}\bigr>.
\end{eqnarray}
since $s(i,i)=1$, one can deduce the following recursive expressions,
\begin{eqnarray}
	&&\bigl<M_{+}^{a}M_{-}^{b}\bigr>+\sum_{\substack{i,j\geq0\\[1pt]i,j\neq a,b}} s(a,i)s(b,j)\bigl<M_{+}^{i}M_{-}^{j}\bigr>
	= \varepsilon_{+}^{a}\varepsilon_{-}^{b}\bigl<N_{+}^{a}N_{-}^{b}\bigr>+\sum_{\substack{i,j\geq0\\[1pt]i,j\neq a,b}} s(a,i)s(b,j)\bigl<N_{+}^{i}N_{-}^{j}\bigr>, \\
	&&\to \bigl<N_{+}^{a}N_{-}^{b}\bigr>=\frac{\bigl<M_{+}^{a}M_{-}^{b}\bigr>}{\varepsilon_{+}^{a}\varepsilon_{-}^{b}}
	+ \sum_{\substack{i,j\geq0\\[1pt]i,j\neq a,b}}s(a,i)s(b,j)\Biggl(\frac{\bigl<M_{+}^{i}M_{-}^{j}\bigr>}{\varepsilon_{+}^{a}\varepsilon_{-}^{b}}-\bigl<N_{+}^{i}N_{-}^{j}\bigr>\Biggr),
	\label{eq:factorial_recursion}
\end{eqnarray}
where $i$ and $j$ are increased from $0$ to $a$ or $b$, but they cannot be $a$ or $b$ simultaneously.
By using Eq.(\ref{eq:factorial_recursion}), one can express $\bigl<N_{+}^{a}N_{-}^{b}\bigr>$ in terms of the combination of $\bigl<M_{+}^{i}M_{-}^{j}\bigr>$\cite{Evan}. 
Moreover, one can apply efficiency corrections to the arbitrary order of cumulant using programming language.
%%%%%%%%%%%%%%%%%%%%%%%%%%%%%%%%%%%%%%%%%%%%%%%%%%%%%%%%%%%%%%%%%%%%%
\subsection{Stirling number of the first kind}
Stirling number of the first kind is defined as
\begin{eqnarray}
	s(n,k)=(-1)^{n-k}
	\begin{bmatrix}
		n \\
		k
	\end{bmatrix},
\end{eqnarray}
where
\begin{eqnarray}
	\begin{bmatrix}
		n \\
		k
	\end{bmatrix}
	=|s(n,k)|, 
\end{eqnarray}
is the unsigned Stirling number of the first kind, which can be calculated by recurrence relation
\begin{eqnarray}
	\begin{bmatrix}
		n+1 \\
		k
	\end{bmatrix}
	=n
	\begin{bmatrix}
		n \\
		k
	\end{bmatrix}
	+
	\begin{bmatrix}
		n \\
		k-1
	\end{bmatrix},
\end{eqnarray}
with the initial conditions
\begin{eqnarray}
	\begin{bmatrix}
		0 \\
		0
	\end{bmatrix}
	=1,~
	\begin{bmatrix}
		n \\
		0
	\end{bmatrix}
	=
	\begin{bmatrix}
		0 \\
		n
	\end{bmatrix}
	=0.
\end{eqnarray}
%\fi
%%%%%%%%%%%%%%%%%%%%%%%%%%%%%%%%%%%%%%%%%%%%%%%%%%%%%%%%%%%%%%%%%%%%%
\subsection{$\Delta K_{3}$\label{app_DK3}}
Efficiency corrections of the third order cumulant are expressed as below:
\begin{eqnarray}
	&&K_{3,{\rm sep}}=\mu_{3}-3\mu_{1}\mu_{2}+2\mu_{1}^{3} \nonumber \\
		       &&=\bigl<N_{+}^{3}\bigr>-3\bigl<N_{+}^{2}N_{-}\bigr>+3\bigl<N_{+}N_{-}^{2}\bigr>-\bigl<N_{-}^{3}\bigr>
			  -3\Bigl(\bigl<N_{+}\bigr>-\bigl<N_{-}\bigr>\Bigr)\Bigl(\bigl<N_{+}^{2}\bigr>-2\bigl<N_{+}N_{-}\bigr>+\bigl<N_{-}^{2}\bigr>\Bigr)+2\Bigl(\bigl<N_{+}\bigr>-\bigl<N_{-}\bigr>\Bigr)^{3} \nonumber \\
		      &&=\frac{1}{\varepsilon_{+}^{3}}\biggl[\bigl<M_{+}^{3}\bigr>+2\bigl<M_{+}\bigr>-3\bigl<M_{+}^{2}\bigr>-3\bigl<M_{+}\bigr>\Bigl(\bigl<M_{+}^{2}\bigr>-\bigl<M_{+}\bigr>\Bigr)+2\bigl<M_{+}\bigr>^{3}\biggr] \nonumber \\
		      &&-\frac{1}{\varepsilon_{-}^{3}}\biggl[\bigl<M_{-}^{3}\bigr>+2\bigl<M_{-}\bigr>-3\bigl<M_{-}^{2}\bigr>-3\bigl<M_{-}\bigr>\Bigl(\bigl<M_{-}^{2}\bigr>-\bigl<M_{-}\bigr>\Bigr)+2\bigl<M_{-}\bigr>^{3}\biggr] \nonumber \\
		      &&-\frac{3}{\varepsilon_{+}^{2}\varepsilon_{-}}\Biggl[\bigl<M_{+}^{2}M_{-}\bigr>-\bigl<M_{+}M_{-}\bigr>-\bigl<M_{-}\bigr>\Bigl(\bigl<M_{+}^{2}\bigr>-\bigl<M_{+}\bigr>\Bigr)-2\bigl<M_{+}\bigr>\bigl<M_{+}M_{-}\bigr>+2\bigl<M_{+}\bigr>^{2}\bigl<M_{-}\bigr>\Biggr] \nonumber \\
			&&+\frac{3}{\varepsilon_{-}^{2}\varepsilon_{+}}\Biggl[\bigl<M_{-}^{2}M_{+}\bigr>-\bigl<M_{-}M_{+}\bigr>-\bigl<M_{+}\bigr>\Bigl(\bigl<M_{-}^{2}\bigr>-\bigl<M_{-}\bigr>\Bigr)-2\bigl<M_{-}\bigr>\bigl<M_{-}M_{+}\bigr>+2\bigl<M_{-}\bigr>^{2}\bigl<M_{+}\bigr>\Biggr] \nonumber \\
		      &&+\frac{3}{\varepsilon_{+}^{2}}\Bigl[\bigl<M_{+}^{2}\bigr>-\bigl<M_{+}\bigr>-\bigl<M_{+}\bigr>^{2}\Bigr]-\frac{3}{\varepsilon_{-}^{2}}\Bigl[\bigl<M_{-}^{2}\bigr>-\bigl<M_{-}\bigr>-\bigl<M_{-}\bigr>^{2}\Bigr]+\frac{\bigl<M_{+}\bigr>}{\varepsilon_{+}}-\frac{\bigl<M_{-}\bigr>}{\varepsilon_{-}}. \label{eq:3rdsep}
\end{eqnarray}
\begin{eqnarray}
	&&K_{3,{\rm ave}}=\frac{1}{\varepsilon^{3}}\biggl[\bigl<M_{+}^{3}\bigr>+2\bigl<M_{+}\bigr>-3\bigl<M_{+}^{2}\bigr>-3\bigl<M_{+}\bigr>\Bigl(\bigl<M_{+}^{2}\bigr>-\bigl<M_{+}\bigr>\Bigr)+2\bigl<M_{+}\bigr>^{3}\biggr] \nonumber \\
		      &&-\frac{1}{\varepsilon^{3}}\biggl[\bigl<M_{-}^{3}\bigr>+2\bigl<M_{-}\bigr>-3\bigl<M_{-}^{2}\bigr>-3\bigl<M_{-}\bigr>\Bigl(\bigl<M_{-}^{2}\bigr>-\bigl<M_{-}\bigr>\Bigr)+2\bigl<M_{-}\bigr>^{3}\biggr] \nonumber \\
		      &&-\frac{3}{\varepsilon^{3}}\Biggl[\bigl<M_{+}^{2}M_{-}\bigr>-\bigl<M_{-}\bigr>\bigl<M_{+}^{2}\bigr>-2\bigl<M_{+}\bigr>\bigl<M_{+}M_{-}\bigr>+2\bigl<M_{+}\bigr>^{2}\bigl<M_{-}\bigr>\Biggr] \nonumber \\
			&&+\frac{3}{\varepsilon^{3}}\Biggl[\bigl<M_{-}^{2}M_{+}\bigr>-\bigl<M_{+}\bigr>\bigl<M_{-}^{2}\bigr>-2\bigl<M_{-}\bigr>\bigl<M_{-}M_{+}\bigr>+2\bigl<M_{-}\bigr>^{2}\bigl<M_{+}\bigr>\Biggr] \nonumber \\
		      &&+\frac{3}{\varepsilon^{2}}\Bigl[\bigl<M_{+}^{2}\bigr>-\bigl<M_{+}\bigr>-\bigl<M_{+}\bigr>^{2}\Bigr]-\frac{3}{\varepsilon^{2}}\Bigl[\bigl<M_{-}^{2}\bigr>-\bigl<M_{-}\bigr>-\bigl<M_{-}\bigr>^{2}\Bigr]+\frac{\bigl<M_{+}\bigr>}{\varepsilon}-\frac{\bigl<M_{-}\bigr>}{\varepsilon}.
\end{eqnarray}
Similarly to $\Delta K_{1}$ and $\Delta K_{2}$ (see Eqs.(\ref{eq:delta_K1}) and (\ref{eq:delta_K2})), we can obtain $\Delta K_{3}$ as
\begin{eqnarray}
	\Delta K_{3}&=&K_{3,{\rm ave}}-K_{3,{\rm sep}} \nonumber \\
		    &\approx&\Biggl[\frac{1}{\varepsilon^{4}}\Bigl[3\bigl(A_{+}+A_{-}\bigr)-\bigl(B_{+}+B_{-}\bigr)\Bigr]+\frac{2}{\varepsilon^{3}}\bigl(C_{+}+C_{-}\bigr)+\frac{2}{\varepsilon^{2}}\bigl(D_{+}+D_{-}\bigr)\Biggr]\Delta\varepsilon,
		\label{eq:delta_K3}
\end{eqnarray}
where constant terms are defined as
\begin{eqnarray*}
	A_{\pm}&=&\bigl<M_{\pm}^{3}\bigr>+2\bigl<M_{\pm}\bigr>-3\bigl<M_{\pm}^{2}\bigr>-3\bigl<M_{\pm}\bigr>\Bigl(\bigl<M_{\pm}^{2}\bigr>-\bigl<M_{\pm}\bigr>\Bigr)+2\bigl<M_{\pm}\bigr>^{3}, \\
	B_{\pm}&=&3\bigl<M_{\pm}^{2}M_{\mp}\bigr>-3\bigl<M_{\pm}M_{\mp}\bigr>-3\bigl<M_{\mp}\bigr>\Bigl(\bigl<M_{\pm}^{2}\bigr>-\bigl<M_{\pm}\bigr>\Bigr)-6\bigl<M_{\pm}\bigr>\bigl<M_{\pm}M_{\mp}\bigr>+6\bigl<M_{\pm}\bigr>^{2}\bigl<M_{\mp}\bigr>, \\
	C_{\pm}&=&3\Bigl(\bigl<M_{\pm}^{2}\bigr>-\bigl<M_{\pm}\bigr>^{2}-\bigl<M_{\pm}\bigr>\Bigr), \\
	D_{\pm}&=&\bigl<M_{\pm}\bigr>.
\end{eqnarray*}
%%%%%%%%%%%%%%%%%%%%%%%%%%%%%%%%%%%%%%%%%%%%%%%%%%%%%%%%%%%%%%%%%%%%%
\subsection{Examples of general expression\label{app_GE}}
In case of $K_{1}$ (see Eq.~(\ref{eq:delta_K1})), we obtain
\begin{eqnarray}
f(M_{+},M_{-})&=&\bigl<M_{+}\bigr>,\qquad\ \ f(M_{-},M_{+})=\bigl<M_{-}\bigr>, \nonumber \\
F[(M_{+},\varepsilon_{+}),(M_{-},\varepsilon_{+})]&=&\frac{\bigl<M_{+}\bigr>}{\varepsilon_{+}},\qquad\ F[(M_{-},\varepsilon_{-}),(M_{+}\varepsilon_{+})]=\frac{\bigl<M_{-}\bigr>}{\varepsilon_{-}}, \nonumber \\
G(M_{+},M_{-},\varepsilon,\Delta\varepsilon)&=&\frac{\Delta\varepsilon}{\varepsilon^2}\bigl<M_{+}\bigr>,\quad G(M_{-},M_{+},\varepsilon,\Delta\varepsilon)=\frac{\Delta\varepsilon}{\varepsilon^2}\bigl<M_{-}\bigr>.  
\end{eqnarray}
Similarly for $K_{2}$ (see Eq.~(\ref{eq:delta_K2})),
\begin{eqnarray}
f(M_{+},M_{-})&=&\bigl<M_{+}^2\bigr>-\bigl<M_{+}M_{-}\bigr>-\bigl<M_{+}^2\bigr>+\bigl<M_{+}\bigr>\bigl<M_{-}\bigr>,\nonumber \\
f(M_{-},M_{+})&=&\bigl<M_{-}^2\bigr>-\bigl<M_{-}M_{+}\bigr>-\bigl<M_{-}^2\bigr>+\bigl<M_{-}\bigr>\bigl<M_{+}\bigr>,\nonumber \\
F[(M_{+},\varepsilon_{+}),(M_{-},\varepsilon_{-})]&=&\frac{\bigl<M_{+}^{2}\bigr>}{\varepsilon_{+}^{2}}-\frac{\bigl<M_{+}\bigr>}{\varepsilon_{+}^{2}}+\frac{\bigl<M_{+}\bigr>}{\varepsilon_{+}}-\frac{\bigl<M_{+}M_{-}\bigr>}{\varepsilon_{+}\varepsilon_{-}}-\frac{\bigl<M_{+}\bigr>^{2}}{\varepsilon_{+}^{2}}+\frac{\bigl<M_{+}\bigr>\bigl<M_{-}\bigr>}{\varepsilon_{+}\varepsilon_{-}},   \nonumber\\ 
F[(M_{-},\varepsilon_{-}),(M_{+},\varepsilon_{+})]&=&\frac{\bigl<M_{-}^{2}\bigr>}{\varepsilon_{-}^{2}}-\frac{\bigl<M_{-}\bigr>}{\varepsilon_{-}^{2}}+\frac{\bigl<M_{-}\bigr>}{\varepsilon_{-}}-\frac{\bigl<M_{-}M_{+}\bigr>}{\varepsilon_{-}\varepsilon_{+}}-\frac{\bigl<M_{-}\bigr>^{2}}{\varepsilon_{-}^{2}}+\frac{\bigl<M_{-}\bigr>\bigl<M_{+}\bigr>}{\varepsilon_{-}\varepsilon_{+}}, \nonumber  \\ 
%G(M_{+},M_{-},\varepsilon,\Delta\varepsilon)&=&-\frac{2\Delta\varepsilon}{\varepsilon^{2}}\Biggl[\frac{\bigl<M_{+}\bigr>-{A}_{+}^{2}}{\varepsilon}-\frac{1}{2}\bigl<M_{+}\bigr>\Biggr], \nonumber\\
G(M_{+},M_{-},\varepsilon,\Delta\varepsilon)&=&-\frac{2\Delta\varepsilon}{\varepsilon^{2}}\Biggl[\frac{X_{+}}{\varepsilon}-\frac{1}{2}\bigl<M_{+}\bigr>\Biggr], \nonumber\\
G(M_{-},M_{+},\varepsilon,\Delta\varepsilon)&=&-\frac{2\Delta\varepsilon}{\varepsilon^{2}}\Biggl[\frac{X_{-}}{\varepsilon}-\frac{1}{2}\bigl<M_{-}\bigr>\Biggr].
\end{eqnarray}

\end{widetext}

% The \nocite command causes all entries in a bibliography to be printed out
% whether or not they are actually referenced in the text. This is appropriate
% for the sample file to show the different styles of references, but authors
% most likely will not want to use it.
\nocite{*}

\bibliography{main}% Produces the bibliography via BibTeX.

\end{document}